\pdfoutput=1

\documentclass[11pt]{article}

\usepackage{authblk}

\usepackage[]{emnlp2021}

\usepackage{times}
\usepackage{latexsym} 

\usepackage[T1]{fontenc}

\usepackage[utf8]{inputenc}

\usepackage{microtype}

\title{\sliceast: Enhancing Code Summarization with Hierarchical Splitting and Reconstruction of Abstract Syntax Trees}

\usepackage{enumitem}
\usepackage{makecell}
\usepackage{xspace}
\usepackage{amsmath,amssymb,amsfonts}
\usepackage{graphicx}
\usepackage{textcomp}
\usepackage{subfigure}
\usepackage{tcolorbox}
\usepackage{booktabs}
\usepackage{tabularx}
\usepackage{lipsum}
\usepackage{multirow}

\newcommand{\myauthornote}[3]{}

\newcommand\yanlin[1]{\myauthornote{yl}{magenta}{#1}}  
\newcommand\enshi[1]{\myauthornote{enshi}{teal}{#1}}

\newcommand\hy[1]{\myauthornote{HY}{red}{#1}}  
\newcommand{\later}[1]{}

\newcommand{\astattgru}{Astattgru\xspace}
\newcommand{\attgru}{Attgru\xspace}
\newcommand{\hdeepcom}{HDeepcom\xspace}
\newcommand{\codenn}{CodeNN\xspace}

\newcommand{\hybriddrl}{HybridDrl\xspace}

\newcommand{\codetoseq}{Code2seq\xspace}

\newcommand{\astnn}{ASTNN\xspace}
\newcommand{\codeastnn}{CodeAstnn\xspace}

\newcommand{\acl}{NCS\xspace}

\newcommand{\sliceast}{CAST\xspace}
\newcommand{\sliceastWithoutCopy}{CAST$_C$}
\newcommand{\sliceastWithoutagg}{CAST$_A$}

\newcommand{\Fig}{Fig.\xspace}
\newcommand{\Tab}{Table\xspace}
\newcommand{\Sec}{Sec.\xspace}
\newcommand{\Eq}{Eq.\xspace}

\newcommand{\Ch}{{\it Ch}}

\usepackage{CJKutf8}
\usepackage[utf8]{inputenc} 

\usepackage{listings}
\usepackage{xcolor}
\definecolor{light-gray}{gray}{0.99}
\usepackage{fancybox}


\newcommand{\figmargin}{0pt}
\newcommand{\secmargin}{0pt}
\newcommand\equalauthorfootnote[1]{%
  \begingroup
  \renewcommand\thefootnote{}\footnote{\textsuperscript{\dag}#1}%
  \addtocounter{footnote}{-1}%
  \endgroup
}

\newcommand\corrauthorfootnote[1]{%
  \begingroup
  \renewcommand\thefootnote{}\footnote{\textsuperscript{\S}#1}%
  \addtocounter{footnote}{-1}%
  \endgroup
}

\author{
Ensheng Shi\textsuperscript{a,\dag}
Yanlin Wang\textsuperscript{b,\dag,\S}
Lun Du\textsuperscript{b}
Hongyu Zhang\textsuperscript{c}\
\\
\textbf{Shi Han}\textsuperscript{b}
\textbf{Dongmei Zhang}\textsuperscript{b}
\textbf{Hongbin Sun}\textsuperscript{a,\S}
\\
\textsuperscript{a}Xi'an Jiaotong University \quad
\textsuperscript{b}Microsoft Research \quad
\textsuperscript{c}The University of Newcastle
\\
{\tt s1530129650@stu.xjtu.edu.cn, hsun@mail.xjtu.edu.cn}\\
{\tt \{yanlwang, lun.du, shihan, dongmeiz\}@microsoft.com } \\
{\tt hongyu.zhang@newcastle.edu.au}\\

}
\begin{document}

\maketitle


\vspace{10pt}
\begin{abstract}

Code summarization aims to generate concise natural language descriptions of source code, which can help improve program comprehension and maintenance. Recent studies show that syntactic and structural information extracted from abstract syntax trees (ASTs) is conducive to summary generation. 
However, existing approaches fail to fully capture the rich information in ASTs because of the large size/depth of ASTs.
In this paper, we propose a novel model \sliceast that hierarchically splits and reconstructs ASTs. 
First, we hierarchically split a large AST into a set of subtrees and utilize a recursive neural network to encode the subtrees. 
Then, we aggregate the embeddings of subtrees by reconstructing the split ASTs to get the representation of the complete AST. 
Finally, AST representation, together with source code embedding obtained by a vanilla code token encoder, is used for code summarization. 
Extensive experiments, including the ablation study and the human evaluation, on benchmarks have demonstrated the power of CAST.
To facilitate reproducibility, our code and data are available at \url{https://github.com/DeepSoftwareAnalytics/CAST}.

\end{abstract}


\section{Introduction}
Code summaries\equalauthorfootnote{The first two authors contribute equally.}\corrauthorfootnote{Yanlin Wang and Hongbin Sun are the corresponding authors.} are concise natural language descriptions of source code 
and they are important for program comprehension and software maintenance. However, it remains a labor-intensive and time-consuming task for developers to document code with good summaries manually. 
\begin{figure*}[t]
    \flushleft
    \subfigure[Source code snippet]{
    \begin{minipage}{0.53\linewidth}
    \flushleft
    \includegraphics[width=\linewidth]{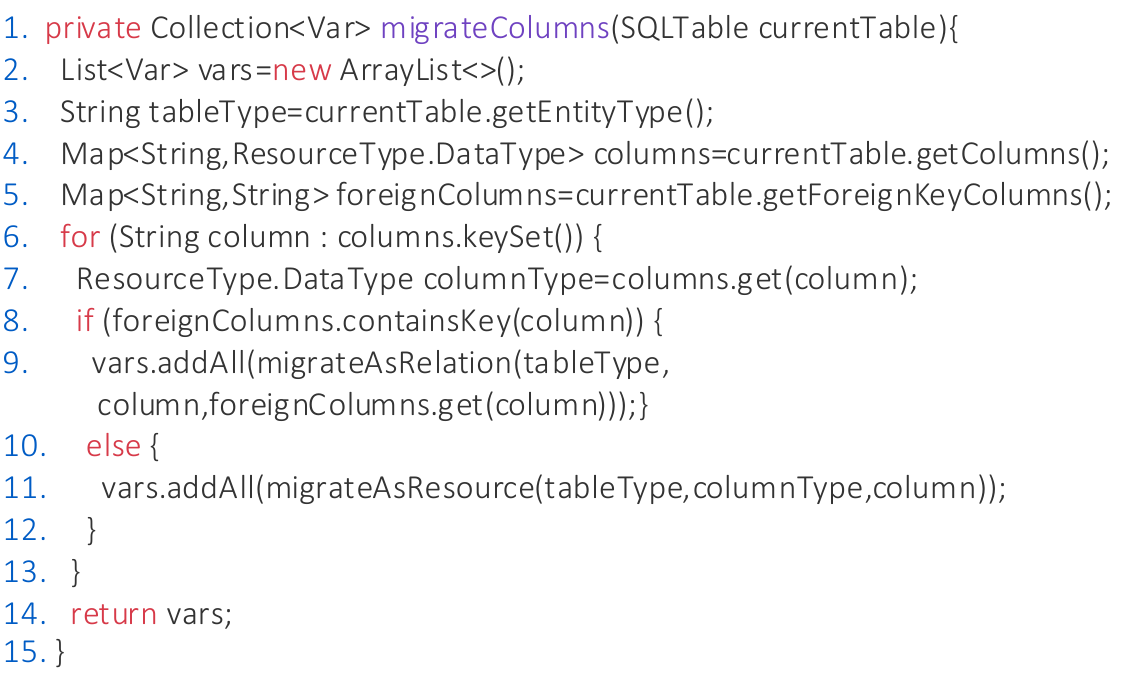}
    \vspace{-10pt}
    \label{fig:ast_slicing_ex:code}
    \end{minipage} 
    }
    \subfigure[Full AST]{
    \begin{minipage}{0.44\linewidth}
        \flushright
        \includegraphics[width=1.0\linewidth]{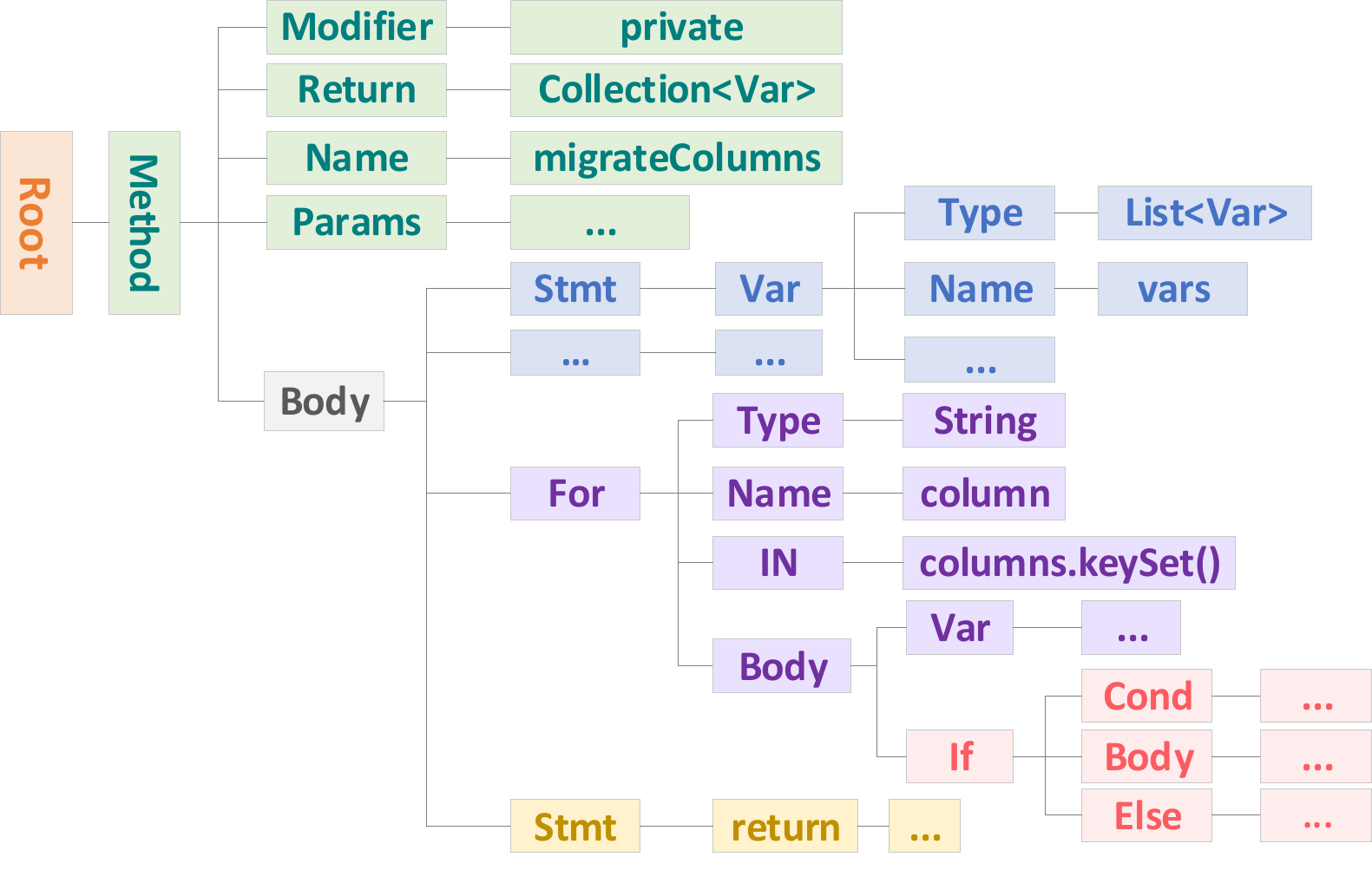}
        \vspace{5pt}
    \end{minipage}%
    \label{fig:ast_slicing_ex:full_ast}
    }%
  \vspace{-1pt}
    \subfigure[Split subtrees]{
    \begin{minipage}{1.0\linewidth}
        \flushleft
          \includegraphics[width=1.0\linewidth]{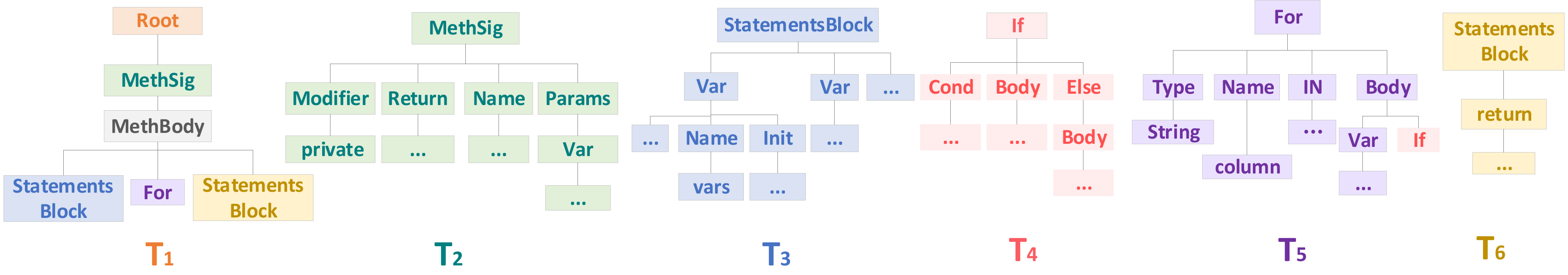}
    \end{minipage}%
    \label{fig:ast_slicing_ex:sliced AST}
    }%
    
    \subfigure[Structure tree]{
    \begin{minipage}{0.14\linewidth}
        \centering
        \includegraphics[width=\textwidth]{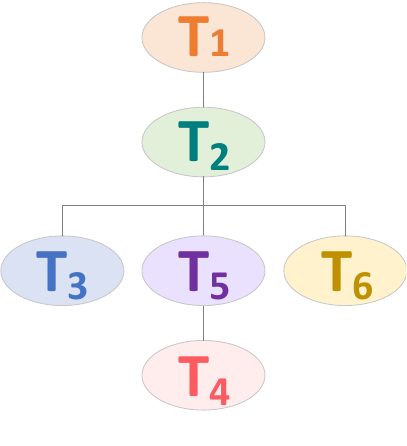}
    \end{minipage}%
    \label{fig:ast_slicing_ex:structure_tree}
    }
    \subfigure[Summaries generated by various approaches.The first raw is the summaries wrote by human. Two sub-sentences in the reference summary are marked in different color.]{
    \begin{minipage}{0.8\linewidth}
      \centering
    \label{fig:ast_slicing_ex:sum}
    \flushright
        \includegraphics[width=\linewidth]{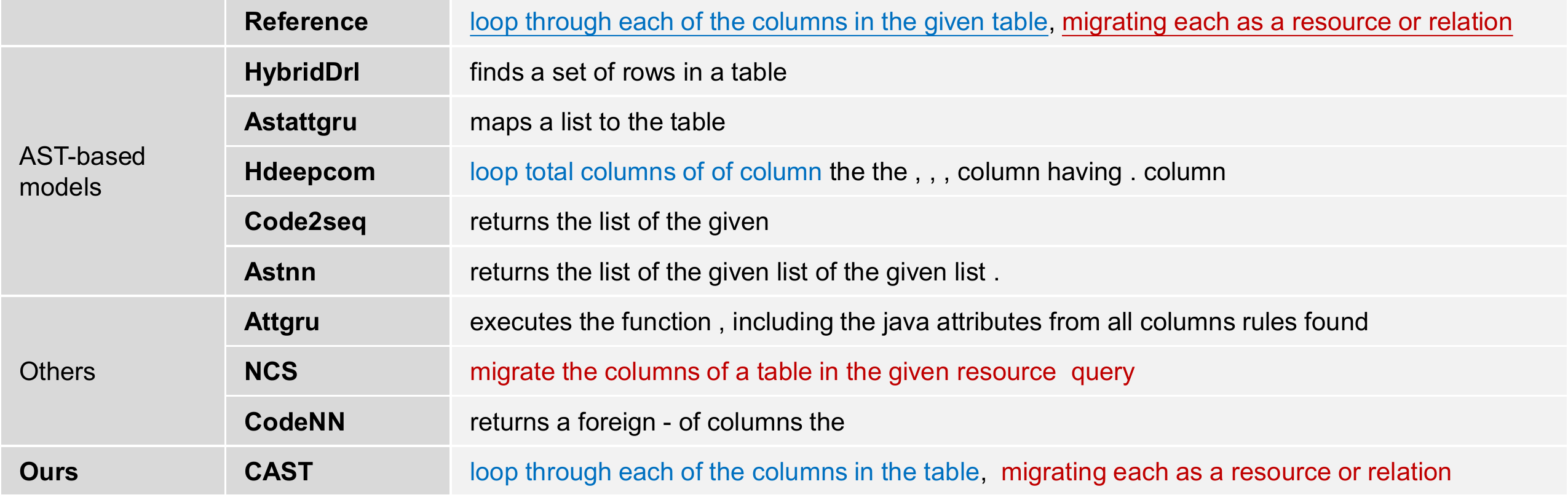}
    \end{minipage}%
    }%
\vspace{-5pt}
\caption{A running example of code, AST, and generated summaries.}
\label{fig:ast_slicing_ex}
\end{figure*}

Over the years, many code summarization methods have been proposed to automatically summarize program subroutines. Traditional approaches such as rule-based and information retrieval-based approaches regard source code as plain
text~\cite{HaiducAM10,HaiducAMM10} without considering the complex grammar rules and syntactic structures exhibited in source code.
Recently, abstract syntax trees (ASTs), which carry the syntax and structure information of code, are widely used to enhance code summarization techniques. 
For example, \citet{HuLXLJ18} propose the structure-based traversal (SBT) method to flatten ASTs and use LSTM to encode the SBT sequences into vectors. 
\citet{hu2019deep} and \citet{LeClairJM19} extend this idea by
separating the code and AST into two input channels, demonstrating the effectiveness of leveraging AST information.
\citet{AlonBLY19,AlonZLY19} extract paths from an AST and represent a given code snippet as a set of sampled paths. Other works~\cite{WanZYXY0Y18,ZhangWZ0WL19,MouLZWJ16} use
tree-based models such as Tree-LSTM, Recursive Neural Network (RvNN), and
Tree-based CNN to model ASTs and improve code summarization. 


We have identified some limitations of the existing AST-based approaches, which lead to a slow training process and/or the loss of AST structural information. 
We now use an example shown in \Fig~\ref{fig:ast_slicing_ex} 
to illustrate the limitations: 
\begin{itemize} 
    \item Models that directly encode ASTs with tree-based neural networks suffer from long training time. \hybriddrl~\cite{WanZYXY0Y18} spends 21 hours each epoch on Funcom~\cite{LeClairJM19}. This is because ASTs are usually large and deep due to the complexity of programs, especially when there are nested program structures. For example, our statistics show that the maximal node number/depth of ASTs of methods in TL-CodeSum~\cite{HuLXLLJ18} and Funcom are 6,165/74 and 550/32, respectively. 
    Moreover, \hybriddrl transforms ASTs into binary trees, leading to deeper trees and more loss of structural information. As shown in \Fig~\ref{fig:ast_slicing_ex:sum}, the main semantics of the code in \Fig~\ref{fig:ast_slicing_ex:code} are not fully captured by \hybriddrl.
    
    \item Linearization methods that flatten ASTs into sequences~\cite{HuLXLJ18,AlonBLY19,AlonZLY19}, by nature, lose the hierarchical information of ASTs. \astnn~\cite{ZhangWZ0WL19} splits an AST into small statement trees to reduce the difficulty of large tree training. However, each subtree contains one statement only and subtrees are later linearized and fed into an RNN, also leading to the loss of hierarchical information.
    From \Fig~\ref{fig:ast_slicing_ex:sum}, we can see that linearization methods \codetoseq~\cite{AlonBLY19}, \astattgru~\cite{LeClairJM19} and \astnn~\cite{ZhangWZ0WL19} fail to capture the main semantics, and \hdeepcom~\cite{hu2019deep} captures only partial semantics.
\end{itemize}
To overcome the above limitations, we propose a novel model \sliceast{} (\underline{C}ode summarization with hierarchical splitting and reconstruction of \underline{A}bstract \underline{S}yntax \underline{T}rees).
The key idea of our approach is to split an AST (\Fig~\ref{fig:ast_slicing_ex:full_ast}) into a set of subtrees (\Fig~\ref{fig:ast_slicing_ex:sliced AST}) at a proper granularity and learn the representation of the complete AST by aggregating its subtrees' representation learned using tree-based neural models. 
First, we split a full AST in a hierarchical way using 
a set of carefully designed rules. 
Second, we use a tree-based neural model RvNN to learn each subtree's representation.
Third, we reconstruct the split ASTs and combine all subtrees' representation by another RvNN to capture the full tree's structural and semantic information. 
Finally, the representation of the complete tree, together with source code embedding obtained by a vanilla code token encoder, is fed to a Transformer decoder to generate descriptive summaries.
\enshi{revised}
Take \Fig~\ref{fig:ast_slicing_ex:code} for example again: there are two sub-sentences in the reference summary. 
The \texttt{For} block (Lines 6, 7 and 13 in \Fig~\ref{fig:ast_slicing_ex:code}) corresponds to the first sub-sentence \emph{``loop through each of the columns in the given table''}, and the \texttt{If} block (Line 8-12) corresponds to the second sub-sentence \emph{``migrating each as a resource or relation''}. 
The semantics of each block can be easily captured when the large and complex AST is split into five subtrees as shown in \Fig~\ref{fig:ast_slicing_ex:sliced AST}. After splitting, $T_5$ corresponds to first sub-sentence and $T_4$ corresponds to the second sub-sentence. When we reconstruct the split ASTs according to \Fig~\ref{fig:ast_slicing_ex:structure_tree}, it is easier for our approach to generate the summary with more comprehensive semantics.

Our method \sliceast has two-sided advantages: 
(1) Tree splitting reduces AST to a proper size to allow effective and affordable training of tree-based neural models.
(2) Different from previous work, we not only split trees but also reconstruct the complete AST using split ASTs. This way, high-level hierarchical information of ASTs can be retained.

We conduct experiments on TL-CodeSum~\cite{HuLXLLJ18} and Funcom \cite{LeClairJM19} datasets, and compare 
\sliceast with the state-of-the-art methods. The results show that our model
outperforms the previous methods in four widely-used metrics Bleu-4, Rouge-L,
Meteor and Cider, and significantly decreases the training time compared to \hybriddrl.
We summarize the main contributions of this paper as follows:

\begin{itemize}[leftmargin=10pt]

\item We propose a novel AST representation learning method based on hierarchical tree splitting and reconstruction. The splitting rule specification and the tool implementation are provided for other researchers to use in AST relevant tasks. 

\item We design a new code summarization approach \sliceast, which incorporates the proposed AST representations and code token embeddings for generating code summaries.
\item We perform extensive experiments, including the ablation study and the human evaluation, on \sliceast and state-of-the-art methods. The results demonstrate the power of \sliceast.
\end{itemize}

\section{Related Work}\label{sec:rw}
\vspace{\secmargin}

\subsection{Source Code Representation}
\vspace{\secmargin}

Previous work suggests various representations of source code for follow-up analysis. 
Allamanis et al.~\shortcite{AllamanisBBS15} and Iyer et al.~\shortcite{IyerKCZ16} consider source code as plain text and use traditional token-based methods to capture lexical information. 
Gu et al.~\shortcite{GuZZK16} use the Seq2Seq model 
to learn intermediate vector representations of queries in natural language to predict relevant API sequences. 
Mou et al.~\shortcite{MouLZWJ16} propose a 
tree-based convolutional neural network to learn program representations.
Alon et al.~\shortcite{AlonZLY19,AlonBLY19} represent a code snippet as a set of compositional paths in the abstract syntax tree. 
Zhang et al.~\shortcite{ZhangWZ0WL19} propose an AST-based Neural Network (ASTNN) that splits each large AST into a sequence of small statement trees and encodes them to vectors by capturing the lexical and syntactical knowledge. 
Shin et al.~\shortcite{ShinABP19} represent idioms as AST segments using probabilistic tree substitution grammars for two tasks: idiom mining and code generation. 
\cite{leclair2020improved,wang2021code} utilize GNNs to model ASTs. There are also works that utilize ensemble model~\cite{du2021single} or pre-trained models~\cite{feng2020codebert,guo2021graphcodebert,bui2021infercode} to model source code.

\subsection{Source Code Summarization}
\label{section:src_sum}
Apart from the works mentioned above, researchers have proposed many approaches to source code summarization over the years. 
For example, 
Allamanis et al.~\shortcite{AllamanisBBS15} create the neural logbilinear context model for suggesting method and class names by embedding them in a high dimensional
continuous space. Allamanis et al.~\shortcite{AllamanisPS16} also suggest a convolutional model for the summary generation that uses attention over a sliding window of tokens. They
summarize code snippets into extreme, descriptive function name-like
summaries. 

Neural Machine Translation based models are also widely used for code summarization~\cite{IyerKCZ16,haije2016automatic,HuLXLJ18,HuLXLLJ18,WanZYXY0Y18,hu2019deep,LeClairJM19,AhmadCRC20,YuHWF020}. 
\textbf{\codenn}~\cite{IyerKCZ16} is the first neural approach for code summarization. It is a classical encoder-decoder framework that encodes code to context vectors with an attention mechanism and then generates summaries in the decoder. \textbf{\acl{}}~\cite{AhmadCRC20} models code using Transformer to capture the long-range dependencies. \textbf{\hybriddrl}~\cite{WanZYXY0Y18} uses hybrid code representations (with ASTs) and deep reinforcement learning. It encodes the sequential and structural content of code by LSTMs and tree-based
LSTMs and uses a hybrid attention layer to get an integrated representation. 
\textbf{\hdeepcom}~\cite{hu2019deep}, \textbf{\astattgru}, and
\textbf{\attgru}~\cite{LeClairJM19} are essentially encoder-decoder network
using RNNs with attention. \astattgru and \hdeepcom utilize a multi-encoder
neural model that encodes both code and AST. \textbf{\codetoseq{}}~\cite{AlonBLY19} represents a code snippet as a
set of AST paths and uses attention to select the relevant paths while
decoding. 
When using neural networks to represent large and deep ASTs, the above work will encounter problems such as gradient vanishing and slow training. \sliceast can alleviate these problems by introducing a more efficient AST representation to generate better code summarie. 



\vspace{\secmargin}
\section{CAST: Code Summarization with AST Splitting and Reconstruction}~\label{sec:mode}

\begin{figure}[t]
\centering
\includegraphics[width=1.0\linewidth]{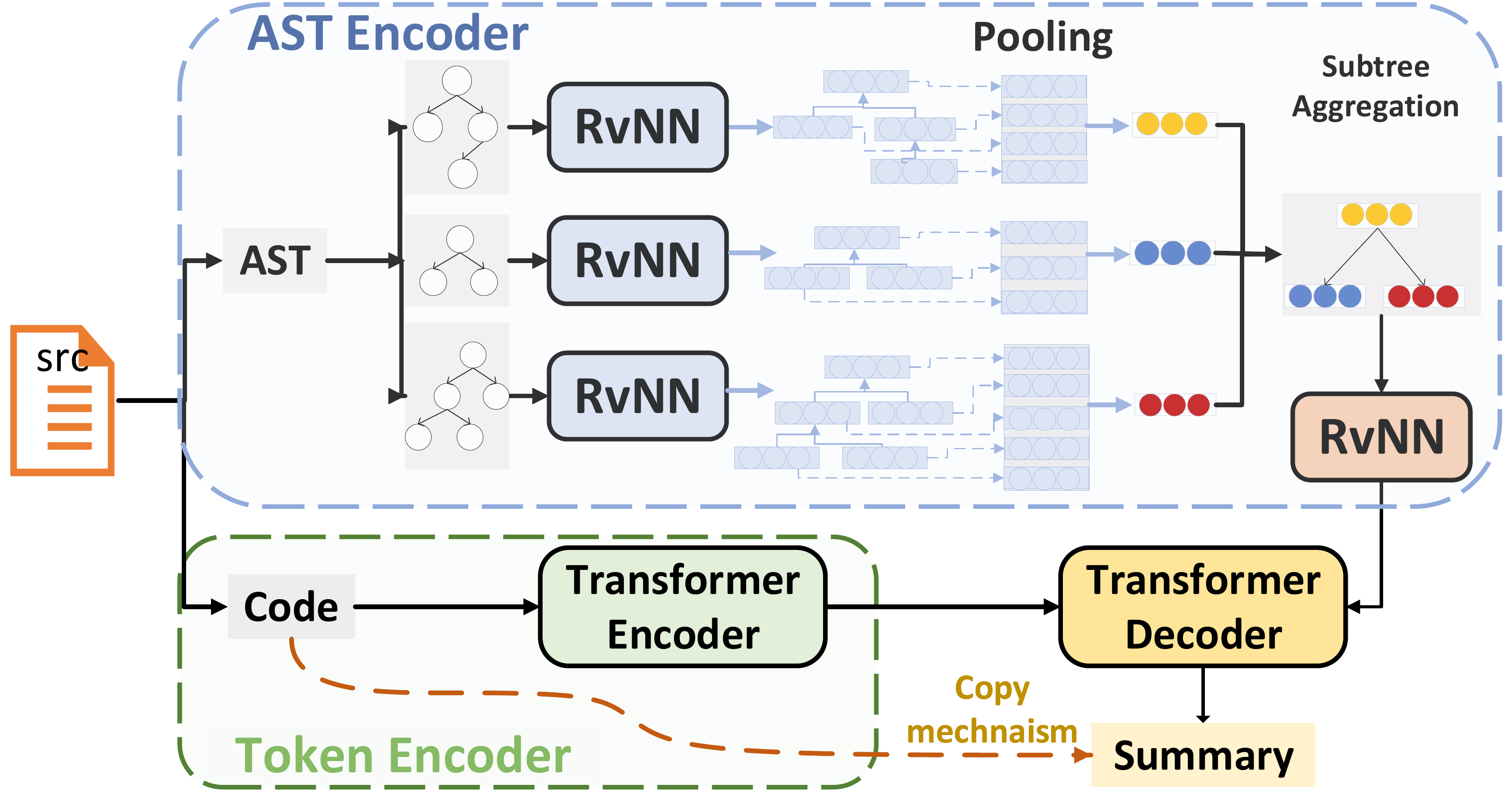}
\caption{\sliceast model structure.}
\label{fig:model}
\end{figure}

This section presents the details of our model. The architecture of 
\sliceast (\Fig~\ref{fig:model}) follows the general Seq2Seq framework and includes three major components: an AST encoder, a code token encoder, and a summary decoder. Given an input method, the AST encoder captures the semantic and structural information of its AST. 
The code token encoder encodes the lexical information of the method. 
The decoder integrates the multi-channel representations from the two encoders and incorporates a copy mechanism~\cite{SeeLM17} to generate the code summary.

\vspace{\secmargin}
\subsection{AST Encoder}

\subsubsection{AST Splitting and Reconstruction}
\label{sec:split_ast}

\begin{figure}[t]
\centering
\includegraphics[width=1.0\linewidth]{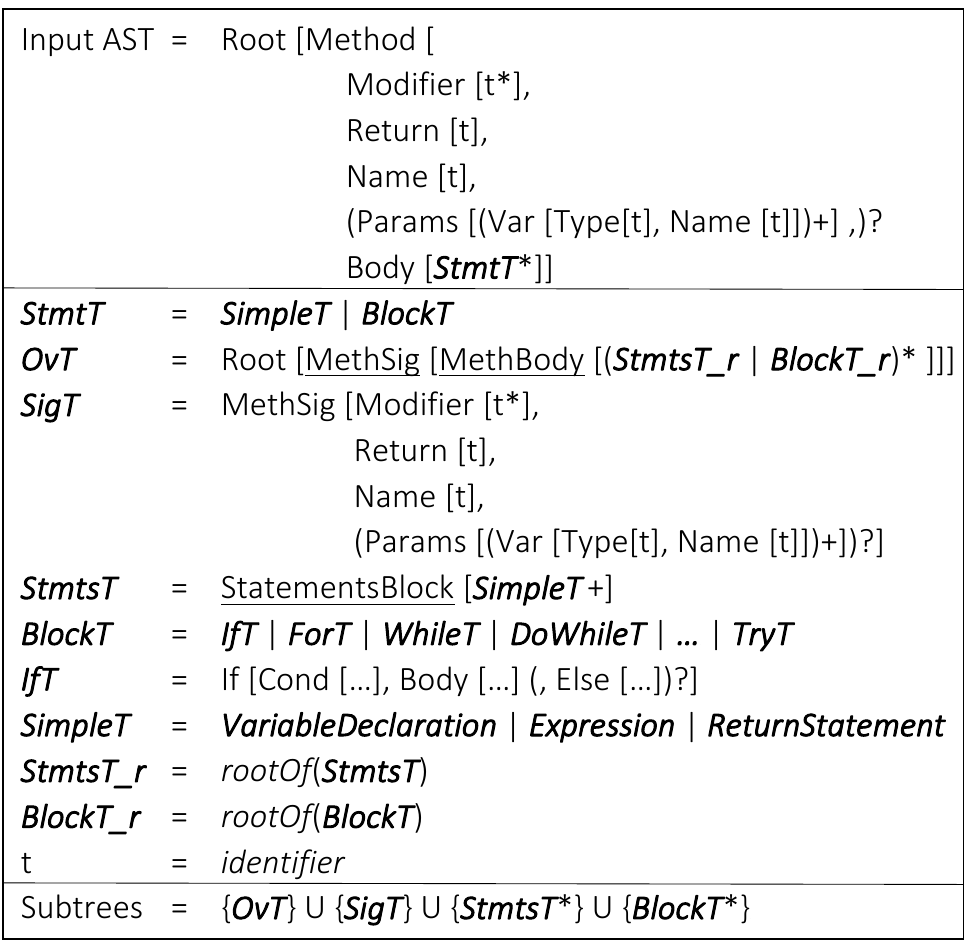}
\caption{AST splitting rule specification. Input is a full \texttt{AST}
and output is the \texttt{Subtrees} set. We use brackets to denote
subtree relation instead of the visualized trees to make the rules compact.  
\textbf{\textit{Bold and italic}} stands for components that can be further
reduced. \underline{Underline} stands for nodes created in the splitting process.}
\label{fig:rule}
\vspace{\figmargin}
\end{figure}

Given a code fragment, we build its AST and visit it by preorder traversal. Each time a composite structure (i.e. \texttt{If}, \texttt{While}, etc.) is encountered, a placeholder node is inserted. The subtree rooted at this statement is split out to form the next level tree, whose semantics will be finally stuffed back to the placeholder. In this way, a large AST is decomposed into a set of small subtrees with the composite structures retained. 

Before presenting the formal tree splitting rules, we provide an illustrative example
in \Fig~\ref{fig:ast_slicing_ex}. The full AST\footnote{The full AST is
omitted due to space limit, it can be found in Appendix.}
(\Fig~\ref{fig:ast_slicing_ex}(b)) of the given code snippet
(\Fig~\ref{fig:ast_slicing_ex}(a)) is split to six subtrees $T_1$ to $T_6$ in
\Fig~\ref{fig:ast_slicing_ex}(c). $T_1$ is the overview tree with non-terminal
nodes \texttt{Root}, \texttt{MethSig}, \texttt{MethBody}, and three terminal nodes
\texttt{StatementsBlock} (blue), \texttt{For}, and \texttt{StatementsBlock} (yellow) corresponding to the three main
segments with \texttt{Line2-5}, \texttt{Line6-13}, and \texttt{Line14} in
\Fig~\ref{fig:ast_slicing_ex}(a), respectively.  
The \texttt{StatementsBlock} (blue) node corresponds to $T_3$ which contains $4$ initialization statements. The \texttt{For} node corresponds to $T_5$  and the \texttt{StatementsBlock} (yellow) node corresponds to $T_6$ which consists of a return statement. Note that each subtree reveals one-level abstraction, meaning that nested structures are abstracted out. Therefore, the \texttt{If} statement nested in
the \texttt{For} loop is split out to the subtree $T_4$, leaving a placeholder \texttt{If} node in $T_5$.

We give the formal definition\footnote{We only present the top-down skeleton and partial rules due to space limitation. The full set of rules, the splitting algorithm, and tool implementation are provided in Appendix.}
 of subtrees in \Fig~\ref{fig:rule}. The goal is
to split a given AST to the subtree set $Subtrees$. In general, all subtrees
are generated by mapping reduction rules on the input AST (similar to mapping
language grammar rules to a token sequence) and the output $Subtrees$ collects four kinds of subtrees: $OvT$, $SigT$, $StmtsT$, and $BlockT$.
$OvT$ is the method overview tree, providing the big picture of the method and
$SigT$ gives the method signature information. To avoid too many scattered
simple statements, we combine sequential statements to form a statements'
block $StmtsT$. We drill down each terminal node in $OvT$ to a subtree $T$
being a $StmtT$ or $BlockT$, providing detailed semantics of nodes in the
overview tree. In the same way, subtrees corresponding to nested structures
(such as \texttt{For} or \texttt{If}) will be split out to form new subtrees.
We split out nested blocks one level at a time until there is no nested block.
 Finally,  we obtain a set of these block-level subtrees. Also, a
\textit{structure tree} (e.g., \Fig~\ref{fig:ast_slicing_ex:structure_tree})
that represents the ancestor-descendant relationships between the subtrees is maintained.
\subsubsection{AST Encoding}
\label{sec:ASTEncoder}
We design a two-phase AST encoder module according to the characteristics of subtrees.
In the first phase, a tree-based Recursive Neural Network (RvNN) followed by a max-pooling layer is applied to encode each subtree. In the second phase, we use another RvNN with different parameters to model the hierarchical relationship among the subtrees.



A subtree $T_t$ is defined as $(V_t, E_t)$ where $V_t$ is the node set and
$E_t$ is the edge set. The forward propagation of RvNN to encode the subtree $T_t$ is formulated as:
\begin{equation}
\footnotesize
\label{equ:recursive-nn}
    \mathbf{h_i^{(t)}} = {\rm tanh} \left(\mathbf{W^C}  \mathbf{c_i^{(t)}} + \frac{1}{\lvert \Ch^t(v_i) \rvert} \sum_{v_j \in \Ch^t(v_i)} \mathbf{W^A} \mathbf{h_j^{(t)}} \right),
\end{equation}

\noindent where $\mathbf{W^C}$ and $\mathbf{W^A}$ are learnable weight
matrices, $\mathbf{h_i^{(t)}}$, $\mathbf{c_i^{(t)}}$, $\Ch(v_i)$ are the hidden state, token embedding, and child set of the node $v_i$, respectively.
Particularly, $\mathbf{h^{(t)}_i}$ equals to
$\mathbf{W^C}  \mathbf{c_i^{(t)}}$ for the leaf node $v_i$.

Intuitively, this
computation is the procedure where each node in the AST aggregates information
from its children nodes. After this bottom-up aggregation, each node has its
corresponding hidden states. Finally, the hidden states of all nodes are aggregated to a vector $\mathbf{s_t}$ through dimension-wise
max-pooling operation, which will be used as the embedding for the whole
subtree $T_t$:
\begin{equation}
\footnotesize
\label{equation:subtree_vectors}
\mathbf{s_t} = {\rm maxpooling}\left(\cup \mathbf{h_i^{(t)}}\right), \forall v_i \in V_t.
\end{equation}



After obtaining the embeddings of all subtrees, we further encode the
descendant relationships among the subtrees. These relationships are
represented in the \textit{structure tree} (e.g., \Fig~\ref{fig:ast_slicing_ex:structure_tree}) $T$, 
thus we apply another RvNN model on $T$:
\begin{equation}
\footnotesize
    \mathbf{h_t^{(a)}} = {\rm tanh} ( \mathbf{W^S}\mathbf{s_t} + \frac{1}{\lvert \Ch(v_t) \rvert} \sum_{v_k \in \Ch(v_t)} \mathbf{W^B}\mathbf{h_k^{(a)}} ).
\end{equation}



There are two main advantages of our AST encoder design. First, it enhances
the ability to capture semantic information in multiple subtrees of a program
by the first layer RvNN, because the tree splitting technique leads to
subtrees that contain semantic information from different modules. In addition, to obtain more important features of the node vectors, we sample all nodes through max pooling. 
The second
layer RvNN can further aggregate information of subtrees according to their relative positions in the hierarchy. 
Second, tree sizes are
decreased significantly after splitting, thus the gradient vanishing and
explosion problems are alleviated. Also, after tree splitting, the depth of
each subtree is well controlled, leading to more stable model training.

\begin{figure}[t]
\centering
\includegraphics[width=0.9\linewidth]{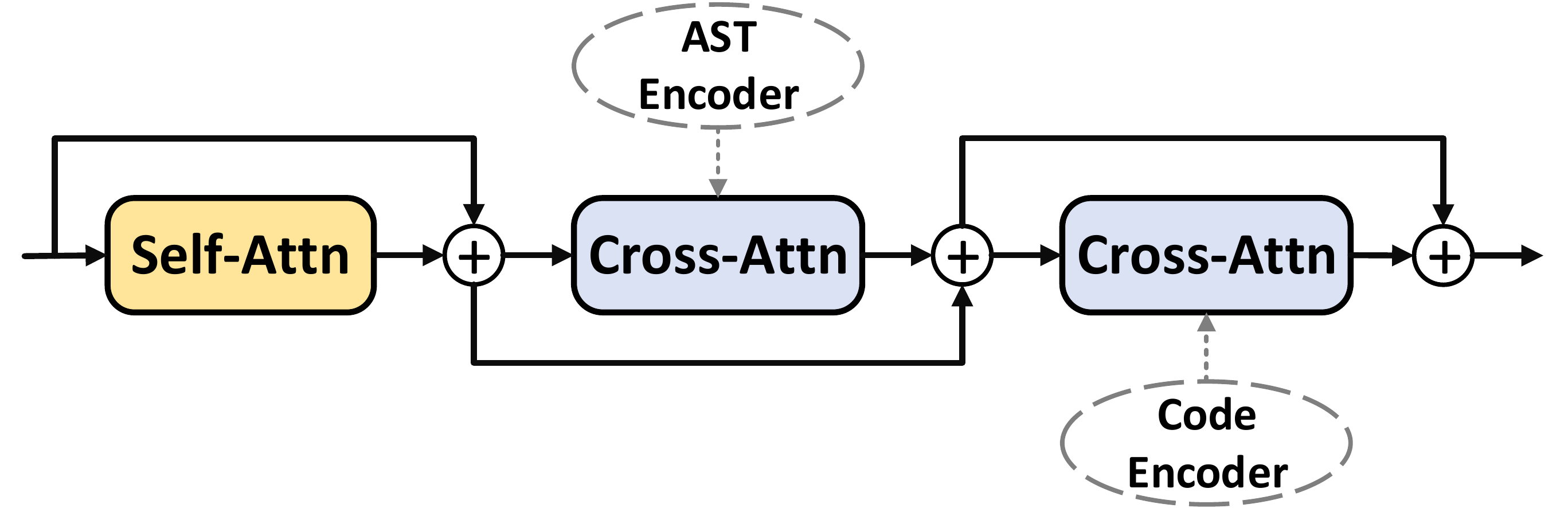}
\caption{The serial strategy for integrating two encoding sources in the decoder.}
\label{fig:decoder}
\end{figure}

\subsection{Code Token Encoder}
\vspace{\secmargin}

The code snippets are the raw data source to provide lexical information for
the code summarization task. Following \cite{AhmadCRC20}, we adopt the code
token encoder using Transformer that is composed of a multi-head self-attention module and a relative position embedding module.
In each attention head, the sequence of code token embeddings $c= (c_1,...,c_n)$ are transformed into output vector $o= (o_1,...,o_n)$
\begin{equation}
\scriptsize
\begin{aligned}
o_{i}&=\sum_{j=1}^{n} \alpha_{i j}\left( \mathbf{W^{V}}c_{j}+a_{i j}^{V}\right), e_{i j}=\frac{(\mathbf{W^{Q}}c_{i}) ^{T}\left( \mathbf{W^{K}}c_{j}+a_{i j}^{K}\right)}{\sqrt{d_{k}}}
\end{aligned}
\end{equation}
where $\alpha_{i j}=\frac{\exp e_{i j}}{\sum_{k=1}^{n} \exp e_{i k}}$, $\mathbf{W^Q}$, $\mathbf{W^K}$ and $\mathbf{W^V}$ are trainable matrices for queries, keys and values; $d_k$ is the dimension of queries and keys; $a_{i j}^{K}$ and $a_{i j}^{V}$ are relative positional representations for positions $i$ and $j$.



\subsection{Decoder with Copy Mechanism} \label{sec:copy_mechanism}
\vspace{\secmargin}
Similar to the code token encoder, we adopt Transformer as the backbone of the
decoder. Unlike the original decoder module in \cite{VaswaniSPUJGKP17}, we need to integrate two encoding sources from code and AST encoders. The serial strategy \cite{libovicky-etal-2018-input} is adopted, which is to compute the encoder-decoder attention one by one for each input encoder (\Fig~\ref{fig:decoder}). In each cross-attention layer, the encoding of ASTs ($h^{(a)}= (h^{(a)}_1,...,h^{(a)}_l)$ flatted by preorder traversal) or codes ($o= (o_1,...,o_n)$) is queried by the output of the preceding summary self-attention $s= (s_1,...,s_m)$.
\begin{equation}
\small
\begin{aligned}
z_{i}&=\sum_{j=1}^{n} \alpha_{i j}\left( \mathbf{W^{V}}h^{(a)}_{j}\right), \alpha_{i j}=\frac{\exp e^{ast}_{i j}}{\sum_{k=1}^{n} \exp e^{ast}_{i k}} \\
y_{i}&=\sum_{j=1}^{n} \alpha_{i j}\left( \mathbf{W^{V}} o_{j}\right), \alpha_{i j}=\frac{\exp e^{code}_{i j}}{\sum_{k=1}^{n} \exp e^{code}_{i k}} \\e_{i j}^{ast}&=\frac{(\mathbf{W_{d}^{Q}}s_{i} )^{T}\left( \mathbf{W_{d}^{K}}h^{(a)}_{j} \right)}{\sqrt{d_{k}}} ,e_{i j}^{code}=\frac{(\mathbf{W_{d}^{Q}}z_{i} )^{T}\left( \mathbf{W_{d}^{K}}o_{j} \right)}{\sqrt{d_{k}}}
\end{aligned}
\end{equation}
where $\mathbf{W_{d}^Q}$, $\mathbf{W_{d}^K}$ and $\mathbf{W_{d}^V}$ are trainable projection matrices for queries, keys and values. $\mathbf{l}$ is the number of subtrees. $\mathbf{m}$ and $\mathbf{n}$ are the length of code and summary tokens, respectively.
Following \cite{VaswaniSPUJGKP17}, we adopt a multi-head attention mechanism in the self-attention and cross-attention layers of the decoder.
After stacking several decoder layers, we add a softmax operator to obtain the generation probability $P^{(g)}_t$ of each summary token.

We further incorporate the copy mechanism~\cite{SeeLM17} 
to enable the decoder to copy rare tokens directly from the input codes. This is motivated by the fact that many tokens (about 28\% in the Funcom dataset) are directly copied from the source code (e.g., function names and variable names) in the summary. Specifically, we learn a copy probability through an attention layer:
\begin{equation}
\footnotesize
    P^{(c)}_t(i) = \frac{{\rm exp}\big(<\mathbf{W^{cp}} \mathbf{h}_i^{(c)},  \mathbf{h}_t^{(s)} >\big)}{\sum_{k=1}^{T_{c}}{\rm exp}\big(<\mathbf{W^{cp}} \mathbf{h}^{(c)}_k,  \mathbf{h}_t^{(s)}>\big)},
\end{equation}
where $P^{(c)}_t(i)$ is the probability for choosing the $i$-th token from source code in the summary position $t$, $\mathbf{h}_i^{(c)}$ is the encoding vector of the $i$-th code token, $\mathbf{h}_t^{(s)}$ is the decoding vector of the $t$-th summary token, $\mathbf{W^{cp}}$ is a learnable projection matrix to map ${h}_i^{(c)}$ to the space of  ${h}_i^{(s)}$, and $T_{c}$ is the code length. 
The final probability for selecting the token $w$ as $t$-th summary token is defined as:
\begin{equation}
\footnotesize 
    P_t(w) = \gamma_t P^{(g)}_t(w) + (1 - \gamma_t) \sum\nolimits_{i:w_i^{(c)} = w} P^{(c)}_t(i),
\end{equation}
where $w_i^{(c)}$ is the $i$-th code token and $\gamma_t$ is a learned combination probability defined as $\gamma_t = {\rm sigmoid} (\Gamma (\mathbf{h}_t^{(s)}))$, 
where $\Gamma$ is a feed forward neural network.
Finally, we use Maximum Likelihood Estimation as the objective function and apply AdamW 
 for optimization.

\section{Experimental Setup}\label{sec:exp}
\vspace{\secmargin}


\subsection{Dataset and Preprocessing}
\label{sec:dataset_pre}
\vspace{\secmargin}



In our experiment, we adopt the public Java datasets TL-CodeSum \cite{HuLXLLJ18} and Funcom \cite{LeClairJM19}, which are widely used in previous studies \cite{AhmadCRC20,HuLXLJ18,hu2019deep,HuLXLLJ18,leclair2020improved,LeClairJM19,zhangretrieval20, WeiLLXJ20}. 
The partitioning of train/validation/test sets follows the original datasets. 
We split code tokens by camel case and snake case, replace numerals and string literals with the generic tokens  \lstinline{<NUM>} and \lstinline{<STRING>}, and set all to lower case. We extract the first sentence of the method's Javadoc description 
as the ground truth summary.
Code that cannot be parsed by the Antlr parser \cite{parr2013definitive} is removed.
At last, we obtain $83,661$ and $2,111,230$ pairs of source code and summaries on TL-CodeSum and Funcom, respectively.

\subsection{Experiment Settings}
\vspace{\secmargin}
We implement our approach based on the open-source project OpenNMT~\cite{klein2017opennmt}.
The vocabulary sizes are $10,000$, $30,000$ and $50,000$ 
for AST, code, and summary, respectively. 
The batch size is set to 128 and the maximum number of epochs is 200/40 for TL-CodeSum and Funcom. For optimizer, we use the AdamW \cite{loshchilov2017decoupled} with the
learning rate $10^{-4}$. To alleviate overfitting, we adopt early stopping
with patience $20$. The experiments are conducted on a server with 4 GPUs
of NVIDIA Tesla V100 and it takes about 10 and 40 minutes each epoch for TL-CodeSum and Funcom, respectively. Detailed hyper-parameter settings and training time can be found in Appendix.





\subsection{Evaluation Metrics}\label{sec:metrics}
\vspace{\secmargin}

Similar to previous work~\cite{IyerKCZ16,WanZYXY0Y18,zhangretrieval20}, we
evaluate the performance of our proposed model based on four widely-used
metrics, including BLEU~\cite{PapineniRWZ02}, Meteor~\cite{BanerjeeL05},
Rouge-L~\cite{lin-2004-rouge} and Cider~\cite{VedantamZP15}. 
These metrics are prevalent metrics in machine translation, text
summarization, and image captioning. Note that we report the scores of
BLEU, Meteor (Met. for short), and Rouge-L (Rouge for short) in percentages since they are in the range of
$[0,1]$. As Cider scores are in the range of $[0,10]$, 
we display them in real values. 
In addition, we notice that the related work on code summarization uses different BLEU implementations, such as BLEU-ncs, BLEU-M2, BLEU-CN, BLEU-FC, etc (named by ~\cite{GrosSDY20}). 
And there are subtle differences in the way the BLEUs are calculated~\cite{GrosSDY20}. 
We choose the widely used BLEU-CN
~\cite{IyerKCZ16,AlonBLY19,FengGTDFGS0LJZ20,wang2020cocogum} as 
the BLEU metric in this work.
Detailed metrics description can be found in Appendix.


\vspace{\secmargin}
\section{Experimental Results}\label{sec:eval}
\begin{table*}[t]
\centering 
\scalebox{0.93}{
\begin{tabular}{cccccccccc}
\toprule
\multirow{2}{*}{Model} & \multicolumn{4}{c}{Funcom} & \multicolumn{4}{c}{TL-CodeSum } \\ 
\cmidrule(r){2-5} \cmidrule(r){6-9} 
                      & BLEU & Met. & Rouge & Cider & BLEU & Met. & Rouge & Cider \\
\midrule
\codenn  &20.93  &11.44 &29.09& 0.90 &22.22&14.08 &33.14 &1.67 \\
\hdeepcom & 25.71   &15.59   &36.07 & 1.42 &23.32 &13.76 &33.94 &1.74  \\
\attgru &27.82  &18.10&39.20&1.84  &29.72&17.03 &38.49 &2.35 \\
\acl &29.18  &19.94 &40.09 &2.15 & 40.63&24.85 &52.00 &3.47  \\
\midrule
\codetoseq   &23.84  &13.84 &33.65 &1.31 &16.09&8.94 &24.21 &0.66\\
\hybriddrl &23.25 &12.55 &32.04 &1.11 &23.51&15.38 &33.86 &1.55 \\
\astattgru &28.17 &18.43&39.56&1.90 &30.78 &17.35 &39.94 &2.31  \\
\codeastnn  &28.27&18.86&40.34&1.94 &41.08&24.95& 51.67&3.49   \\
\midrule
\sliceastWithoutagg{} &30.56  &20.96 &42.46 &2.30 &43.76 &27.15 & 54.09 &3.84\\
\sliceastWithoutCopy{} & 30.35 &20.65 &42.22 &2.24 &43.81 &26.95&53.53&3.82 \\ 
\textbf{\sliceast{}} & \textbf{30.83} &\textbf{20.96} &\textbf{42.71} &\textbf{ 2.31} &\textbf{45.19} & \textbf{27.88} &\textbf{55.08} &\textbf{3.95} \\
\bottomrule
\end{tabular}
}
\caption{Comparison with baselines.} 
\label{tab:cmpsota}
\vspace{\figmargin}
\end{table*}

\subsection{The Effectivness of \sliceast}
\label{sec:RQ1}
\vspace{\secmargin}
We evaluate the effectiveness of \sliceast{} by comparing it to the 
recent DNN-based code summarization models introduced in \Sec~\ref{section:src_sum}: \codenn{}, \hybriddrl{}, \hdeepcom{}, \attgru{}, \astattgru{}, \codetoseq{}, and \acl{}. 
To make a fair comparison, we extend \astnn to \codeastnn, with an additional code token encoder as ours, so that the only difference between them is the AST representation.

From the results in \Tab~\ref{tab:cmpsota}, we can see that \sliceast outperforms all the baselines on both datasets.  
\codenn{}, \codetoseq{}, \attgru{}, and \acl{} only use code or AST information. Among them, \acl{} performs better because it applies a transformer to capture the long-range dependencies among code tokens. \astattgru and \codeastnn outperform \attgru because of the addition of AST channels. 
Note that our model outperforms other baselines 
even without the copy mechanism or aggregation. 
This is because we split an AST into block-level subtrees and each subtree contains relatively complete semantics. On the contrary, related work such as \astnn splits an AST into statement-level subtrees, which only represent a single statement and relatively fragmented semantics. 
\subsection{Comparison of Different AST Representations}\label{sec:RQ2} 
\vspace{\secmargin}

We evaluate the performance of different AST representations by comparing \sliceast with \codetoseq, \hybriddrl, \astattgru, and \codeastnn. 
\Tab~\ref{tab:cmpsota} shows that \sliceast performs the best among them. 
As linearization-based methods,
\astattgru flattens an AST to a sequence and \codetoseq obtains a set of paths
from an AST, both losing some hierarchical information of ASTs naturally. As tree-based methods, \hybriddrl transforms ASTs to binary
trees and trains on the full ASTs with tree-based models. This leads to
AST structural information loss, gradient vanishing problem, 
and slow training process (21 hours each epoch in Funcom)\footnote{See training time details in Appendix \Tab 2 and 3.}. 
Both \codeastnn and \sliceast perform better than \hybriddrl, \codetoseq, and \astattgru because they split a large AST into a set of small subtrees, which can alleviate the gradient vanishing problem. Our \sliceast achieves the best performance and we further explain it from two aspects: splitting granularities of ASTs. and the AST representation learning.

For splitting granularities of ASTs, \codeastnn is statement-level splitting, leading to subtrees 71\% smaller than ours on TL-CodeSum\footnote{See dataset statistics in Appendix \Tab 5 to 8.}.
Therefore, it may not be able to capture the syntactical information and semantic information.
In terms of AST representation learning,  \codeastnn and \sliceast all use RvNN and Max-pooling to learn the representation of subtrees but different ways to aggregate them.
The former applies a RNN-based model to aggregate the subtrees. It only captures the sequential structure and the convergence becomes worse as the number of subtrees increases~\cite{BengioFS93}. The latter 
applies RvNN to aggregate all subtrees together according to
their relative positions in the hierarchy, 
which can combine the semantics of subtrees well.
\subsection{Ablation Study}\label{sec:ablationstudy}
\vspace{\secmargin}
To investigate the usfulness of the subtree aggregation (\Sec~\ref{sec:ASTEncoder}) and the copy mechanism (\Sec~\ref{sec:copy_mechanism}), we conduct ablation studies on two variants of \sliceast. The results of the ablation study are given in the bottom of \Tab~\ref{tab:cmpsota}.
\vspace{\secmargin}
\begin{itemize}[leftmargin=8pt]
    \item \sliceastWithoutagg: \sliceast without subtree aggregation, which directly uses the subtree vectors obtained by \Eq (\ref{equation:subtree_vectors}) 
    as AST representation. Our results show that the performance of \sliceastWithoutagg{} drops compared to \sliceast{} (except for Met. in Funcom), demonstrating that it is beneficial to reconstruct and aggregate information from subtrees. 
    \item \sliceastWithoutCopy: \sliceast without copy mechanism. 
Our results show that \sliceast outperforms \sliceastWithoutCopy{}, confirming that the copy mechanism can copy tokens (especially the out-of-vocabulary ones) from input code to improve the performance of summarization.

\end{itemize}

\vspace{\secmargin}
\subsection{Human Evaluation}
\vspace{\secmargin}

\begin{table}[t]
\centering \footnotesize \setlength{\tabcolsep}{2.8pt}
\begin{tabular}{cccc}
\toprule
Model       &Informativeness        &Naturalness            &Similarity \\ \midrule
\sliceast    &\textbf{2.74}(1.29)   &\textbf{3.08}(1.23)     &\textbf{2.66}(1.29) \\  
\astattgru  & 2.26(1.05)          & 2.46(1.31)            &2.02(1.09) \\
\acl      &2.39(1.10)           &2.78(1.19)            &2.17(1.14) \\
\codeastnn     &2.44(1.08)          &3.00(1.13)           &2.20(1.14)  \\  \bottomrule
\end{tabular}
\caption{Results of human evaluation (standard deviation in parentheses).} 
\label{tab:human_evaluation}
\vspace{\figmargin}
\end{table}

Besides textual similarity based metrics, we also conduct a human evaluation by following the previous work~\cite{IyerKCZ16,Liu0T0L19,hu2019deep,WeiLLXJ20} to evaluate semantic similarity of the summaries generated by \sliceast{}, \astattgru{}, \acl{} and \codeastnn. We randomly choose 50 Java methods from the testing sets (25 from TL-CodeSum and 25 from Funcom) and their summaries generate by four approaches. Specially, we invite 10 volunteers with more than 3 years of software development experience and excellent English ability. Each volunteer is asked to assign scores from 0 to 4 (the higher the better) to the generated summary from the three aspects: \textbf{similarity} of the generated summary and the ground truth summary, \textbf{naturalness} (grammaticality and fluency), and \textbf{informativeness} (the amount of content carried over from the \emph{input code} to the generated summary, ignoring fluency). Each summary is evaluated by four volunteers, and the final score is the average of them. 

\Tab~\ref{tab:human_evaluation} shows that \sliceast{} outperforms others in all three aspects. Our approach is better than other approaches in Informative, which means that our approach tends to generate summaries with comprehensive semantics.
In addition, we confirm the superiority 
of our approach using Wilcoxon signed-rank tests~\cite{wilcoxon1970critical} for the human evaluation. And the results\footnote{See Appendix Table 9} reflect that the improvement of \sliceast{} over other approaches is statistically significant with all p-values smaller than 0.05 at 95\% confidence level (except for \codeastnn on Naturalness).

\vspace{\secmargin}
\section{Threats to Validity}\label{sec:threats}
\vspace{\secmargin}
There are three main threats to validity.
First, we evaluate and compare our work only on a Java dataset, although
in principle, the model should generalize to other languages, experiments are
needed to validate it. Also, AST splitting algorithm need to be implemented
for other languages by implementing a visitor to AST. 


Second, in neural network model design, there are many orthogonal
aspects such as 
different token embeddings, whether to use beam search, teacher forcing.
When showing the generality of \sliceast, we have done the experiments in a
controlled way. A future work might be to do all experiments in a more
controlled way and the performance of \sliceast could rise further when
combined with all other orthogonal techniques. 

Third, summaries in the datasets are collected by
extracting the first sentences of Javadoc. Although this is a common
practice to place a method's summary at the first sentence of Javadoc, there
might still be some mismatch summaries. A higher quality
dataset with better summaries collecting techniques is needed in the future. 

\vspace{\secmargin}
\section{Conclusion}\label{sec:con}
\vspace{\secmargin}
In this paper, we propose a new model \sliceast that splits the AST of
source code into several subtrees, embeds each subtree, and aggregates 
subtrees' information back to form the full AST representation. This
representation, along with code token sequence information, is then fed into a
decoder to generate code summaries. Experimental results have demonstrated the
effectiveness of \sliceast and confirmed the usefulness of the abstraction 
technique. We believe our work sheds some light on future research by pointing
out that there are better ways to represent source code for intelligent code
understanding. 

\section{Acknowledgement}
We thank reviewers for their valuable comments on this work. This research was supported by National Key R\&D Program of China (No. 2017YFA0700800).
We would like to thank Chin-Yew Lin, Jian-Guang Lou, Jiaqi Guo and Qian Liu for their valuable suggestions and feedback during the paper writing process. We also thank the participants of our human evaluation for their time. 

\bibliographystyle{acl_natbib}
\bibliography{ref}

\end{document}